\documentclass[final,english]{bullsrsl}[2022/06/15]

\usepackage[latin1]{inputenc}
\usepackage[T1]{fontenc}
\usepackage[export]{adjustbox}

\usepackage{natbib} 
\usepackage{graphicx}

\begin{document}
\title{Instabilities in models of supergiants MWC 137 and MWC 314}

\author[affil={1}, corresponding]{Sugyan}{Parida}
\author[affil={1}]{ Abhay Pratap}{Yadav}
\author[affil={2}]{Santosh}{Joshi}
\affiliation[1]{Department of Physics and Astronomy, National Institute of Technology, Rourkela - 769008, India }
\affiliation[2]{Aryabhatta Research Institute of Observational Sciences, Manora Peak, Nainital - 263002, India}
\correspondance{522ph6001@nitrkl.ac.in}
\date{10th June 2023}
\maketitle


%

\begin{abstract}
In several B-type supergiants photometric and spectroscopic variabilities together with episodes of 
enhanced mass-loss have been observed.  
Here we present the preliminary results of linear stability analysis followed by 
nonlinear numerical simulations in two B-type supergiants MWC 137 and MWC\,314. 
All the considered models of MWC\,137 having mass in the range of 30 M$_{\odot}$ to 70 M$_{\odot}$ are 
unstable while for the case of MWC\,314 models with mass below 31 M$_{\odot}$ are unstable. 
The instabilities have been followed into nonlinear regime for selected models of these 
two supergiants. During the nonlinear numerical simulations, instabilities lead to finite 
amplitude pulsation with a well defined saturation level in the considered models of MWC\,137 with mass greater than 42 M$_{\odot}$.
The model of MWC\,314 with mass of 40 M$_{\odot}$ - the suggested mass for the primary star - does not show any
instabilities both in linear stability analysis and nonlinear numerical simulations. Velocity amplitude
reaches to 10$^7$ cm/s in the nonlinear regime for the model of MWC\,314 with mass of 30 M$_{\odot}$. 
Further extensive numerical simulations and observations are required to understand the origin of the observed
variabilities in these stars.      

\end{abstract}

\keywords{Massive stars, Supergiants, B-type stars, MWC\,137, MWC\,314, Instabilities in stars}

\section{Introduction}
Massive stars (initial mass $\geq$ 8 M$_\odot$) play a crucial role in chemical enrichment, star formation, and dynamics of galaxies.
These stars have a variety of brief transitional phases during which they show significant changes in their properties such as the 
luminosity, surface temperature and radius. Massive stars are subject to 
significant mass loss  generally through stellar winds or surface eruptions \citep{puls_2008,smith_2014}.
MWC\,137 is a galactic B[e] type star with a reported variability of 1.9 d using the data of the Transiting Exoplanet Survey Satellite (TESS) \citep{Ricker_2015}. The cause of this variability is unknown, however it is suspected to be caused by pulsation \citep{kraus_2021}. MWC\,314 has a galactic supergiant whose evolutionary phase is still a matter of debate \citep[e.g.][]{richardson_2016, frasca_2016}. \citet{frasca_2016} has concluded that MWC\,314 is a binary system with a B[e] type supergiant and an undetected companion. As reported by \citet{richardson_2016}, this binary system exhibits pulsational behaviour as seen by the Microvariability and Oscillations of Stars (MOST) satellite \citep{walker_2003} in two modes with periods of 0.77 and 1.42 d, respectively.

In the present paper, we intend to study instabilities in models of MWC\,137 and the primary star of  MWC\,314. In earlier studies, it has been found that instabilities may be caused by pulsations, surface eruptions and re-arrangement of stellar structure \citep{grott_2005, yadav_2016, yadav_2017}.A comparison with observed variabilities in these two stars requires an extensive linear stability analysis followed by non-linear numerical simulations which will be presented elsewhere.
Section \ref{models} contains the parameters and a description of the models. Section \ref{result} briefly describes the acquired results, followed by a discussion and conclusion in Section \ref{d_and_c}.

\begin{center}

\begin{table}[!ht]
\centering
\begin{minipage}{156mm}
\caption{Luminosity, surface temperature, mass range, and chemical composition of the considered supergiants. In the chemical composition coloumn, X, Y and Z represent the hydrogen mass fraction, helium mass fraction, and mass fraction of heavier elements, respectively.}
\end{minipage}
\label{tabl}
\bigskip

\scalebox{0.95}{
\begin{tabular}{|c|c|c|c|c|}
\hline
Star Name & \begin{tabular}[c]{@{}c@{}}Luminosity \\ {[}log (L/L$_{\odot}$){]}\end{tabular} & \begin{tabular}[c]{@{}c@{}}Surface temperature\\ {[}K{]}\end{tabular} & \begin{tabular}[c]{@{}c@{}}Mass Range\\ {[}M$_{\odot}${]}\end{tabular} & Chemical Composition\\
\hline
MWC\,137    & 5.84 & 28200  & 30 - 70 M$_{\odot}$ & X = 0.70, Y = 0.28, Z = 0.02 \\
\hline
MWC\,314     & 5.7 & 18000   & 25 - 45 M$_{\odot}$ & X = 0.70, Y = 0.28, Z = 0.02 \\
\hline
\end{tabular}
}
\end{table}
\end{center}

\section{Considered Models of MWC 137 and MWC 314} \label{models}

To understand the origin of the variability in MWC\,137 and MWC\,314, we have constructed models of MWC\,137 and MWC\,314 using known parameters such as luminosity, surface temperature and chemical composition mentioned in Table\,\ref{tabl}. For MWC\,137, the values of luminosity and surface temperature are adopted from \citet{kraus_2021}. In the case of MWC\,314, the value of these parameters are taken from \citet{richardson_2016} and \citet{lobel_2013}.The masses of these two stars are not precisely known. We have taken a range of mass 30 to 70 M$_{\odot}$ for MWC\,137 and 25 to 45 M$_{\odot}$ for MWC\,314. The choice of mass range has been taken in such a way that the earlier suggested or determined values of the mass of these stars are in the considered range \citep{kraus_2021,lobel_2013}.

To construct the models of these two massive stars, we have integrated the standard stellar structure equations \citep{2013sse..book.....K} from surface up to a temperature of the order of 10$^7$ K. We have taken Stefan Boltzmann's law and photospheric pressure as boundary conditions. To avoid the complications of nuclear reactions, we have limited our present study to the envelope of these stars. In many massive O and B type stars, earlier studies hare unstableave shown that the instabilities are dominantly present in the envelope \citep[e.g.,][]{epstein_1950, yadav_2016,yadav_2017b}. Magnetic field and rotation have been disregarded. Schwarzschild's criteria is adopted for the onset of convection and OPAL tables are used for opacity \citep{rogers_1992,rogers_1996,iglesias_1996}.



\section{Results} \label{result}
\subsection{Linear stability analysis}
\begin{figure}[!ht]
\centering 
\includegraphics[width = 1.07\textwidth, center]{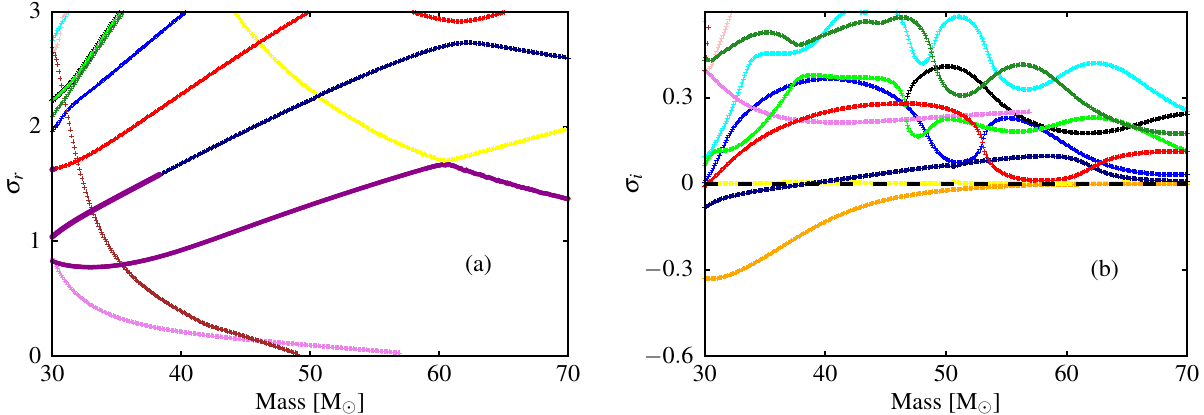}
\bigskip
\begin{minipage}{12cm}
\caption{Real (a) and imaginary (b) parts of eigenfrequencies are plotted as the function of mass in solar unit for the models of MWC\,137. Negative imaginary parts and dark magenta lines in the real parts of eigen frequencies are representing unstable modes.}
\normalsize
\label{modal_1}
\end{minipage}
\end{figure}

\begin{figure}
\centering 
\includegraphics[width = 1.07\textwidth, center]{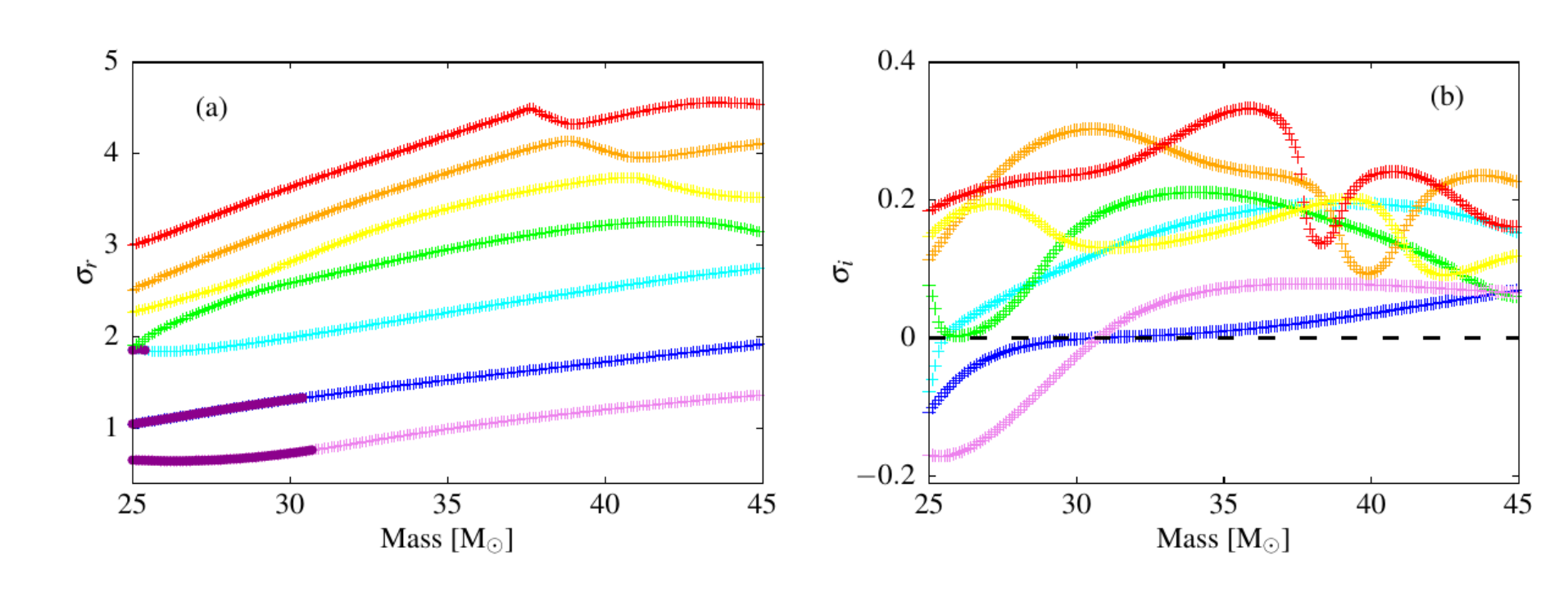}
\bigskip
\begin{minipage}{12cm}
\caption{Same as Fig.\,\ref{modal_1} but for models of  MWC\,314. The models under consideration have three unstable modes.}
\normalsize
\label{modal_2}
\end{minipage}
\end{figure}

Linearized pulsation equations for radial perturbations form a fourth order eigenvalue problem \citep{baker_1962, baker_1965}. This system of equations are solved using the Riccati method as described by \citet{ gautschy_1990b}. The obtained eigenfrequencies are normalised with the global free fall timescale ($\sqrt{R^3/3GM}$ ; where G, M, and R are the gravitational constant, mass, and radius, respectively ).The outcomes of the linear stability analysis are presented in Figs.\,\ref{modal_1} and \ref{modal_2} where the real and imaginary parts of eigenfrequencies are plotted as a function of stellar mass. Diagrams consisting of eigenfrequencies as a function of a stellar parameter (such as stellar mass or surface temperature) are known as modal diagrams \citep{saio_1998, gautschy_1990b}. In the adopted numerical scheme, negative imaginary parts correspond to unstable modes while positive imaginary parts denote damped modes. Further details on the interaction of modes and eigenfrequencies can be found in \citet{gautschy_1990}.

For the case of MWC\,137, the real and imaginary parts of the eigenfrequencies are given as a function of mass in Fig.\,\ref{modal_1}. As the eigenfrequencies are normalised by the global free fall timescale, smaller values of the real part of the eigenfrequencies ($\sigma_r$) correspond to low order modes. Three low order modes ($\sigma_r$ < 2) are unstable in this modal diagram. One mode is unstable in all models with a mass in the range of 30 to 70 M$_{\odot}$. We note the presence of avoided crossings in model having mass of 60.4 M$_{\odot}$ with two modes having $\sigma_r$ nearly equal to 1.5 and 2.8 respectively. The pulsation periods associated with the unstable radial modes are in the range of 1.1 to 3.9 d. The observed period of 1.9 d for the star MWC\,137 is in this range \citep{kraus_2021}. The modal diagram given in Fig.\,\ref{modal_2} for MWC\,314 shows that the models below 31 M$_{\odot}$ all have unstable modes. Two modes are excited in the models having mass in the range of 25 to 31 M$_{\odot}$.The strength of the instabilities increases for models with a higher luminosity-to-mass ratio (see Fig \ref{modal_2}) as for a fixed luminosity, this ratio is higher for lower mass models. Apart from these two unstable modes, another mode is excited in models having a mass close to 25 M$_{\odot}$. For the case of MWC\,314, the excited radial modes have pulsation periods in the range of 4.5 to 12.8 d. The periods of observed variabilities (0.77 and 1.42 d) substantially differ from these values.

\subsection{Non-linear numerical simulations}
Linear stability analyses have shown that all the considered models of MWC\,137 are unstable while models below 31 M$_{\odot}$ are unstable for MWC\,314. A linear stability analysis can not predict the final amplitude of the pulsation. To find out the final fate of the unstable models, we have performed non-linear numerical simulations for selected models of  MWC\,137 and MWC\,314. The reader is referred to \citet{grott_2005} for the equations and numerical procedures used here to follow the instability in non-linear regime.

\begin{figure}[!ht]
\centering
\includegraphics[width = 1.07\textwidth, center]{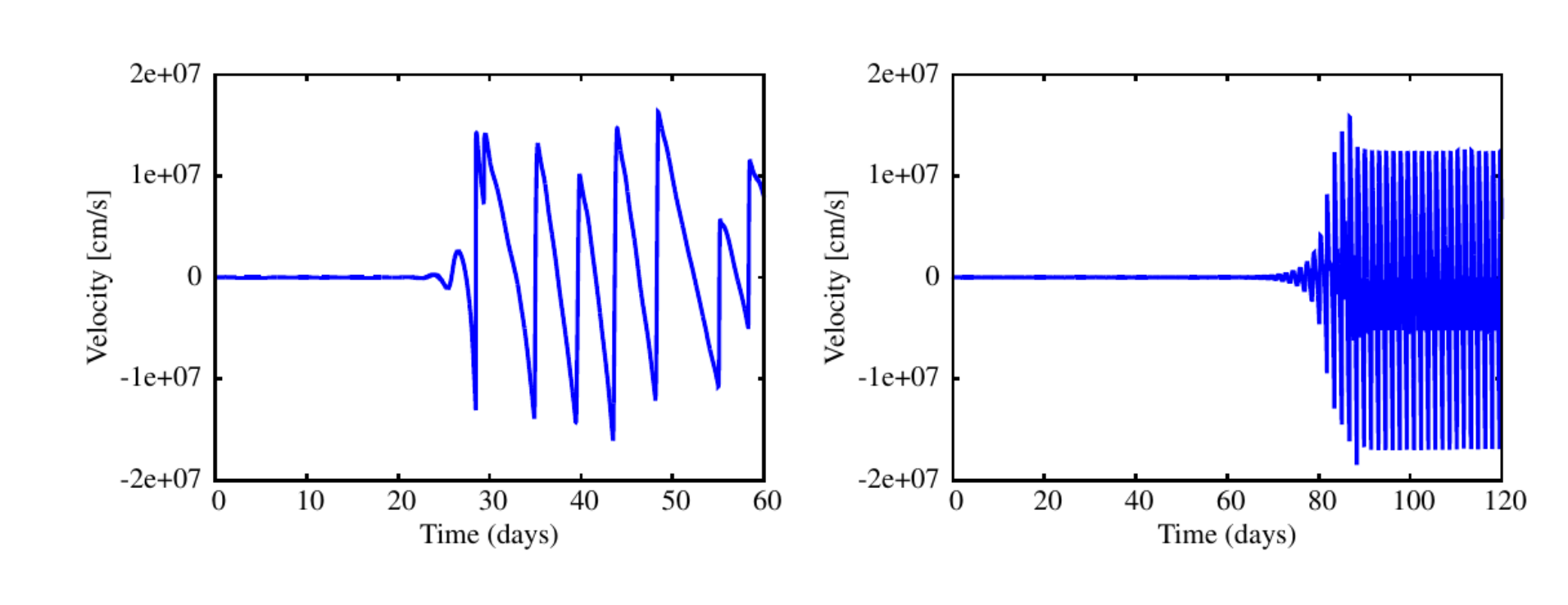}
\bigskip
\begin{minipage}{12cm}
\caption{ Variation of the velocity associated with the outermost grid point during
the nonlinear simulation for the models of MWC\,137 having mass of 37 M$_{\odot}$ (left) and 50 M$_{\odot}$ (right)}.
\normalsize
\label{vel_1}
\end{minipage}
\end{figure}

For MWC\,137, we have considered three models including the mass of 37 M$_\odot$. This mass (37$^{+9} _{-5}$ M$_\odot$) was derived by \citet{kraus_2021}.
The variation in velocity associated with the outmost grid point as a function of time for a model having mass of 37 M$_{\odot}$ and 50 M$_{\odot}$ is given in Fig.\,\ref{vel_1}. The code picks up physical instabilities from the numerical noise without any external perturbation and saturates in the non-linear regime with a velocity amplitude of the order of 1.5 $\times$ 10$^7$ cm/s for 37 M$_{\odot}$ and 1.25 $\times$ 10$^7$ cm/s for 50 M$_{\odot}$. In this preliminary analysis, we note that models having a mass greater than 42 M$_{\odot}$ have a well defined saturation in the velocity amplitude as in the case of 50 M$_{\odot}$ (see Fig. \ref{vel_1}).

\begin{figure}[!ht]
\centering
\includegraphics[width = \textwidth]{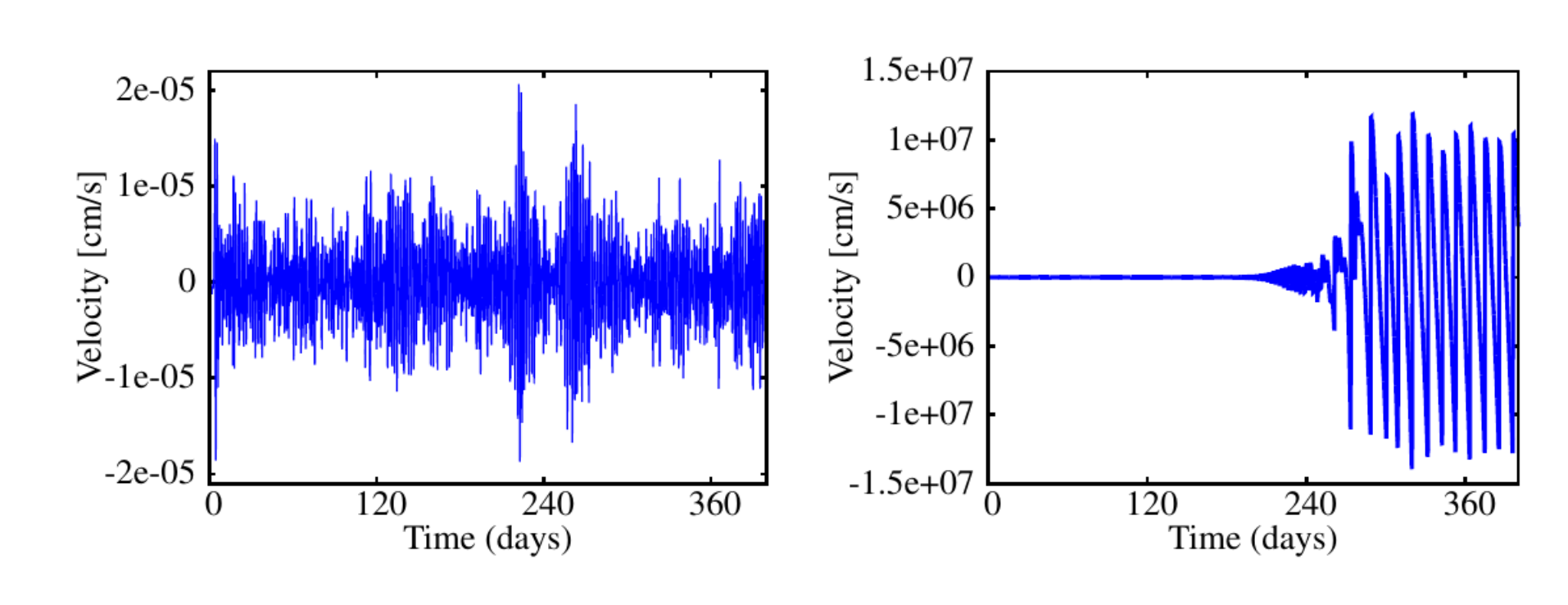}
\bigskip
\begin{minipage}{12cm}
\caption{ Variation of the velocity associated with the outermost grid point during
 the nonlinear simulation for a model of MWC\,314 having mass of 40 M$_{\odot}$ (left)
 and 30 M$_{\odot}$ (right).}
\normalsize
\label{vel_2}
\end{minipage}
\end{figure}

For MWC\,314, models below 31  M$_{\odot}$ are found to be radially unstable in the linear stability analysis. We have considered two models having a mass of 30 and 40 M$_{\odot}$ for non-linear numerical simulation. \citet{lobel_2013} have suggested $\sim$ 40 M$_{\odot}$ as the mass of the primary star. The model of 30  M$_{\odot}$ is considered to examine the final fate instabilities present in unstable models of MWC\,314. The velocity profile associated with the outermost grid point for two models having mass of 30 and 40 M$_{\odot}$ for the primary star of MWC\,314 is given in Fig.\,\ref{vel_2}. In the case of 40 M$_{\odot}$, we note that velocity amplitude remains of the order of numerical noise. We do not find any physical instability for the model of 40 M$_{\odot}$ during non-linear numerical simulation which is in agreement with the outcome of the linear stability analysis as the models of mass greater than 31 M$_{\odot}$ are stable. The model with mass 30 M$_{\odot}$ is unstable and velocity amplitude saturates with a value of approximately 10$^7$ cm/s (see Fig.\,\ref{vel_2}).

\section{Discussion and Conclusion} \label{d_and_c}
We have performed linear stability analyses followed by non-linear numerical simulations in models of MWC\,137 and MWC\,314. Low order radial modes are unstable in models of the considered supergiants. For MWC\,137, excited modes have periods in the range of 1.1 to 3.9\,d. Therefore the observed variability of 1.9\,d in MWC\,137 can be explained by low order radial modes. On the other hand, excited modes in models of MWC\,314 have periods in the range between 4.5 to 12.8\,d. Hence the observed variability of MWC\,314 (0.77 and 1.42\,d) can not be explained by the considered radial modes. A systematic linear stability analysis followed by extensive non-linear numerical simulations are required to find out the origin of these variabilities. Follow-up observations using observational facilities within the Belgo-Indian Network for Astronomy and Astrophysics (BINA) observational facilities can enhance our present understanding about these two massive stars.

\begin{acknowledgments}
Financial support from Science and Engineering Research Board (SERB), India through 
Core Research Grant (CRG/2021/007772) is gratefully acknowledged. We sincerely acknowledge the constructive comments and suggestions of the referee.
\end{acknowledgments}

\begin{furtherinformation}

\begin{orcids}

\orcid{0009-0000-3108-8744}{Sugyan}{Parida}
\orcid{0000-0001-8262-2513}{Abhay Pratap}{Yadav}
\orcid{0009-0007-1545-854X}{Santosh}{Joshi}


\end{orcids}

\begin{authorcontributions}
All authors have contributed significantly.  
\end{authorcontributions}

\begin{conflictsofinterest}
The authors declare no conflict of interest.
\end{conflictsofinterest}

\end{furtherinformation}

\bibliographystyle{bullsrsl-en}

\bibliography{extra}

\end{document}